\documentclass[10pt]{article}
\usepackage{amssymb,amsmath}
\usepackage{graphicx}
\usepackage{color}
\usepackage{hyperref}
\usepackage{curves,multicol}
\usepackage{geometry}
\usepackage{times}
\usepackage{bm}
\usepackage{ulem}
\usepackage{ucs}
\usepackage[intoc]{nomencl}
\usepackage{float}
\usepackage{subfig}
\usepackage{units}
\usepackage{titlesec}
\usepackage{curves,multicol}
\usepackage{mathptmx}
\usepackage{lscape}
\usepackage{comment}
\usepackage[labelfont=bf]{caption}
\usepackage{lipsum}   
\usepackage[none]{hyphenat}
\usepackage{setspace}
\usepackage{changepage}
\usepackage[english]{babel} 
\usepackage{blindtext}
\usepackage[numbers]{natbib}

\makeatletter

\renewcommand\@biblabel[1]{}

\renewcommand{\section}{\@startsection
{section}
{1}
{0mm}
{-\baselineskip}
{0.5\baselineskip}
{\normalfont\bfseries\MakeUppercase}} 

{}

\renewcommand{\subsection}{\@startsection
{subsection}
{2}
{0mm}
{0.5\baselineskip}
{0.25\baselineskip}
{\bfseries\normalsize}} 

\makeatother

\begin{document}
\sloppy

\setcounter{secnumdepth}{-1} 

\vspace*{-1.0cm}
\begin{flushright} \vbox{
34$^{\mathrm{th}}$ Symposium on Naval Hydrodynamics\\
Washington, DC, USA, 26$^{\mathrm{th}}$ June--1$^{\mathrm{st}}$ July 2022}
\end{flushright}

\vskip0.65cm
\begin{center}
\textbf{\LARGE
High-Fidelity Simulation and Novel Data Analysis of the Bubble Creation and Sound Generation Processes in Breaking Waves\\[0.35cm]
}

\Large Q. Gao$^{1,*}$, G. B. Deane$^{2}$, S. Basak$^{1}$, U. Bitencourt$^{1}$, and L. Shen$^{1}$\\ 

($^{1}$Department of Mechanical Engineering and St. Anthony Falls Laboratory, University of Minnesota, Minneapolis, MN 55455, USA, $^{2}$Marine Physical Laboratory, Scripps Institution of Oceanography, University of California at San Diego, La Jolla, CA 92093, USA)\\
\vspace*{0.25cm}

\end{center}

\begin{multicols*}{2}

\section{Abstract}
Recent increases in computing power have enabled the numerical simulation of many complex flow problems that are of practical and strategic interest for naval applications. A noticeable area of advancement is the computation of turbulent, two-phase flows resulting from wave breaking and other multiphase flow processes such as cavitation that can generate underwater sound and entrain bubbles in ship wakes, among other effects. Although advanced flow solvers are sophisticated and are capable of simulating high Reynolds number flows on large numbers of grid points, challenges in data analysis remain. Specifically, there is a critical need to transform highly resolved flow fields described on fine grids at discrete time steps into physically resolved features for which the flow dynamics can be understood and utilized in naval applications. This paper presents our recent efforts in this field. In previous works, we developed a novel algorithm to track bubbles in breaking wave simulations and to interpret their dynamical behavior over time~(Gao et al.,~2021a). We also discovered a new physical mechanism driving bubble production within breaking wave crests~(Gao et al.,~2021b) and developed a model to relate bubble behaviors to underwater sound generation~(Gao et al.,~2021c). In this work, we applied our bubble tracking algorithm to the breaking waves simulations and investigated the bubble trajectories, bubble creation mechanisms, and bubble acoustics based on our previous works.
\section{Introduction}
\label{SECintroduction}
Bubbles play critical roles in ocean--atmosphere processes and naval applications, such as ship hydrodynamics. Specifically, it is important to understand the creation mechanisms, transport behaviors, and acoustic properties of bubbles because they can affect the drag on ship hulls and ships' acoustic and optical signatures. Consequently, numerous experimental and numerical studies have been conducted on the bubbles in breaking waves and their applications to ship hydrodynamics. For example, Deane \& Stokes~(2002), Wang et al.~(2016), Deike et al.~(2016), Gao et al.~(2018), Yu et al.~(2020), and Chan et al.~(2021) investigated the bubble size spectrum in breaking waves or free-surface turbulence and demonstrated that super-Hinze-scale bubbles (bubbles whose radii exceed the Hinze scale) follow a $-10/3$ power-law scaling. Super-Hinze-scale bubbles are created by a fragmentation cascade process~(Garrett et al.,~2000), where large bubbles break up into smaller bubbles due to turbulent fluctuations. Furthermore, Castro et al.~(2014) and Li et al.~(2016) investigated bubble distributions with regard to ship hydrodynamics.

 The above works have advanced our understanding of the physics of bubbly flows in the context of breaking waves and ship hydrodynamics. However, some challenges remain in this field. For example, these works studied the statistics of bubbles at only single instances in time because of the limitation of the conventional bubble identification algorithm, which cannot track bubbles temporally and thus cannot detect their evolutionary behavior over time. Therefore, the fragmentation, coalescence, and trajectories of bubbles were not captured in previous studies. Recently, we developed a novel algorithm (Gao et al.,~2021a) to track bubbles and detect their evolutionary behaviors over time by making connections among bubbles between sequential time instances. Specifically, our algorithm classifies bubble behaviors into five categories. Based on these bubble behaviors, the algorithm derives constraints on the bubble volumes, velocities, and positions at two sequential time instances and then tries to make connections among the bubbles at these successive instances in time by selecting the minimum value within a network of error functions defined by these constraints. By following the networks among bubbles over time, this algorithm is capable of temporally tracking bubbles to obtain their trajectories. Moreover, the algorithm can be used to identify newly formed bubbles when they are not associated with any bubbles at the previous time step. The algorithm here is limited to binary fragmentation and binary coalescence events. In other words, the algorithm cannot detect the event that a bubble breaks up into three and more~(multiple fragmentation) or the event that three or more bubbles merge into one~(multiple coalescence). This is because establishing network becomes much complex when considering multiple fragmentation and coalescence.

To ascertain the mechanism responsible for creating bubbles, previous studies have focused primarily on the bubble fragmentation cascade process. During the bubble fragmentation cascade process, large bubbles are fragmented into smaller ones due to turbulent fluctuation. Recently, Gao et al.~(2021b) studied the bubble generation dynamics in a breaking wave. In their work, they identified a bubble formation mechanism called air cylinder instability and provided a formula expressing its theoretical dispersion. Different from the fragmentation cascade process, the cylinder instability produces bubbles through an interface instability. They compared simulation results with the theory and found good agreement. However, some questions regarding the bubble creation mechanism remain. For example, how many air cylinders are created in a breaking wave, and how many bubbles are created when an air cylinder breakup? Furthermore, how important is the cylinder instability mechanism to the overall formation of bubbles? In the present work, we examine the wave breaking process, bubble creation events, and the bubble size spectrum to quantify the importance of the air cylinder instability to the production of bubbles to some extent.

In addition, the abovementioned bubble tracking and event detection algorithm can be used to identify bubble creation events, which is useful for calculating wave noise. In ocean waves, wave noise is generated by newly formed bubbles. Gao et al.~(2021c) developed a simulation framework for calculating the sound radiated by newly formed bubbles within a breaking wave crest. The framework contains three components: (1)~a two-phase flow solver, (2)~a bubble event detection algorithm, and (3)~a wave noise model for calculating bubble acoustics. In the present work, we apply this framework and investigate the wave noise spectrogram.

The remainder of this paper is organized as follows. First, we describe the numerical method, including the two-phase flow solver, bubble tracking algorithm, and wave noise model. Then, we discuss the bubble entrainment process, bubble size spectrum, bubble trajectories, and bubble acoustics. Finally, some conclusions are drawn.

\section{Numerical Method}
\label{SECmethods}
\subsection{High-fidelity simulation of wave breaking and bubble dynamics}
The simulations are performed by solving the incompressible Navier--Stokes equations on a fixed Eulerian grid, where the air and water phases are treated as a coherent system with varying physical properties. The governing equations are
\begin{equation}
    \label{eq:NS}
    \frac{D\boldsymbol{u}}{D t}=-\frac{1}{\rho} \left(\nabla p+\rho \boldsymbol{g}+\nabla \cdot \left(2\mu\boldsymbol{D}\right)+\sigma\kappa \delta \left(\boldsymbol{x}_{s}\right)\right)
\end{equation}
and
\begin{equation}
    \label{eq:continuous}
    \nabla \cdot \boldsymbol{u}=0,
\end{equation}
where $\boldsymbol{u}$ is the velocity, $\rho$ is the density, $p$ is the pressure, $\boldsymbol{g}$ is the acceleration due to gravity, $\mu$ is the dynamic viscosity, $\boldsymbol{D}\equiv\left(\nabla\boldsymbol{u}+\nabla\boldsymbol{u}^{T}\right)/2$ is the deformation tensor, $\sigma$ is the coefficient of surface tension, $\kappa$ is the interface curvature, $\delta$ is the Dirac delta function, and $\boldsymbol{x}_s$ denotes the position on the air--water interface. The in-house WOW code used here implements the coupled level set and volume of fluid (CLSVOF) method~(Sussman \& Puckett, 2000) to capture the air--water interface. In the level set~(LS) method, $\phi$ is the signed LS distance function, which is positive in water and negative in air and takes a value of $0$ at the air--water interface. In the volume of fluid~(VOF) method, the volume fraction function $F$ is defined as the ratio of the volume of water in a grid cell to the overall grid cell volume. The CLSVOF method utilizes the advantages of both the LS and the VOF methods to precisely describe the interface geometry and ensure the conservation of mass. The governing equations for the LS and volume fraction functions are
\begin{equation}
    \label{eq:level_set}
    \frac{\partial \phi}{\partial t} + \nabla \cdot \left(\boldsymbol{u} \phi \right) = 0
\end{equation}
and
\begin{equation}
    \label{eq:F}
    \frac{\partial F}{\partial t} + \nabla \cdot \left(\boldsymbol{u} F \right) = 0,
\end{equation}
respectively. Using the LS method, the density and viscosity are expressed as
\begin{equation}
    \label{eq:density}
    \rho = H\left(\phi\right) \rho_w + \left(1-H\left(\phi\right)\right) \rho_a
\end{equation}
and
\begin{equation}
    \label{eq:viscosity}
    \mu = H\left(\phi\right) \mu_w + \left(1-H\left(\phi\right)\right) \mu_a,
\end{equation}
where $H$ is a smoothed Heaviside function and the subscripts `$a$' and `$w$' denote air and water, respectively. The solution procedure for the two-phase flow solver consists of three key steps: (i)~updating the fluid density and viscosity using equations~\eqref{eq:density} and \eqref{eq:viscosity} based on the LS function; (ii)~obtaining the velocity and pressure fields by solving the Navier--Stokes equations~\eqref{eq:NS} and \eqref{eq:continuous}; and (iii)~advancing the LS and volume fraction functions using equations~\eqref{eq:level_set} and \eqref{eq:F}, respectively. More information about the two-phase flow simulation method can be found in Yang et al.~(2018) and Gao et al.~(2021a).
\begin{figure}[H]
	\centering
	\includegraphics[width=\linewidth]{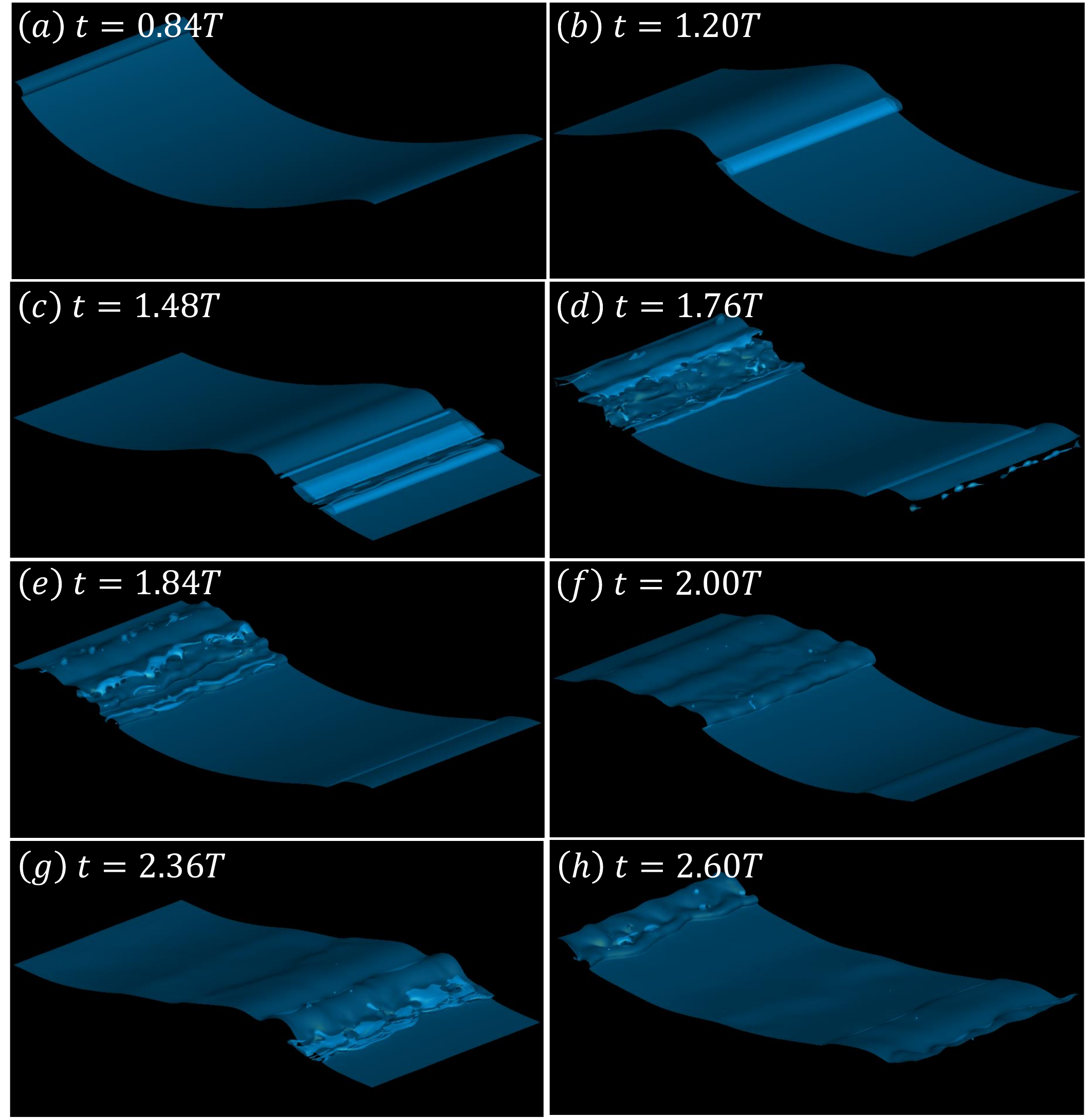}
	\caption{\label{fig:wavesurface38} Visualization of a wave simulation with $S=0.38$ at different times. (a):~A jet forms on the wave crest. (b)--(c):~The overturning wave crest strikes the wave surface, producing multiple cavities and splash jets. (d)--(e):~Cavities break up into bubbles close to the wave surface. (f):~Bubbles are generated near the wave surface and quickly merge with the surface. (g)--(h):~The breaking wave entrains another group of air cylinders and bubbles that quickly merge with the surface.}
\end{figure}
The canonical problem of a Stokes wave with a large initial steepness is analyzed here. The wavelength $\lambda$ is $25\,\mathrm{cm}$. The simulation domain is set to be $\lambda \times \lambda \times \lambda/2$. Periodic boundary conditions are applied in the streamwise and spanwise directions, while a free slip boundary condition is adopted in the vertical direction. Three cases are analyzed with different initial wave steepnesses of $S=0.38$, $0.45$, and $0.55$ corresponding to types of waves from a spilling breaker to a strong plunging breaker. The density ratio $\rho_a/\rho_w$ and viscosity ratio $\mu_a/\mu_w$ are set to $0.0012$ and $0.0154$, respectively. The Reynolds number, Froude number, and Weber number are set as $10^4$, $1$, and $8417.5$, respectively. The simulation time step is $dt = 4 \times 10^{-5}T\approx1.6 \times 10^{-5}\,\mathrm{s}$, where $T$ is the wave period. A uniform-resolution $512\times 512 \times 256$ grid is used, and the grid size is approximately $0.5\,\mathrm{mm}$. The time step for tracking bubble over time or detecting bubble events are selected to be $\Delta t=500 dt \approx 8\times 10^{-3}\mathrm{s}$. Higher-resolution cases will be performed in future studies with an increase in computational power or with more advanced numerical algorithms, for example, adaptive mesh refinement~(Zeng et al.,~2022).

\subsection{Bubble identification and tracking}
The breaking wave surface and bubbles can be illustrated by the zeroth isosurface of the LS function. To identify bubbles from the LS function field and calculate the bubble statistics, the connected component labeling~(CCL) algorithm~(Hermann,~2010; Tomar et al.,~2010; Wang et al.,~2016; Deike et al.,~2016; Bakshi et al.,~2016; Hsiao et al.,~2017; Gao et al.,~2018; Hendrickson et al.,~2019) has been employed in many studies. In the simulations, bubbles comprise grid cells with air surrounded by water. The CCL method uses a breadth-first search~(BFS) algorithm to find and tag the grid cells that contain air and are connected with each other. Although the CCL method has been successfully applied to resolve many physical problems, it can identify bubbles within the flow field at only a single time instance. In contrast, tracking bubbles and detecting their evolutionary behaviors (such as their fragmentation and coalescence) over time, an algorithm whose capabilities exceed those of the conventional CCL method is required.

\begin{figure}[H]
	\centering
	\includegraphics[width=\linewidth]{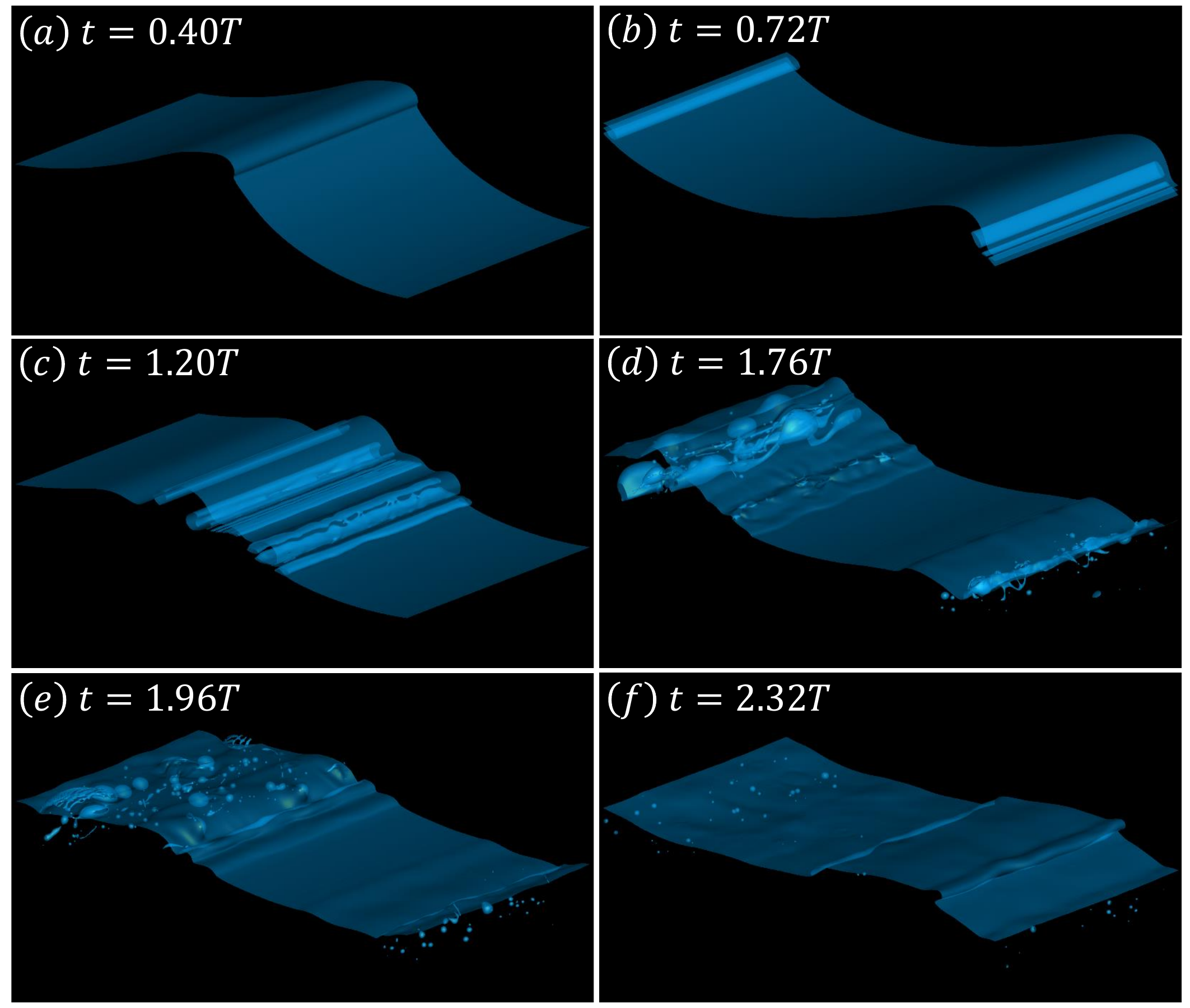}
	\caption{\label{fig:wavesurface45} Visualization of a wave simulation with $S=0.45$ at different times. (a):~A jet forms on the wave crest. (b)--(c):~The overturning wave crest strikes the wave surface, producing multiple cavities and splash jets. (d)--(e):~Large cavities break up into large bubbles due to the instability of the cylinder wall, and larger bubbles fragment into smaller ones in a fragmentation cascade. (f):~Larger bubbles rise toward the wave surface faster than smaller bubbles, leaving the latter suspended underwater for a relatively long time.}
\end{figure}
Here, we track bubbles and detect their behaviors using a novel algorithm called the optimal network~(ON) method, which was proposed in our recent work (Gao et al.,~ 2021a). The ON method classifies bubble behaviors into five categories: continuity, fragmentation, coalescence, formation, and extinction. Continuity means that a bubble is only transported by the flow and that its volume does not change. Fragmentation is the process wherein a larger bubble breaks up into smaller ones, while coalescence is the reverse process of fragmentation; both fragmentation and coalescence are assumed to be binary. Finally, formation and extinction represent a bubble forming and disappearing, respectively, at the wave surface. The ON method tracks bubbles and detects bubble events by making connections among bubbles at sequential time instances. Based on the detected bubble events, we can derive the relationship among bubble volumes, velocities, and positions at two sequential time instances. For example, assuming continuity between two sequential time instances, two bubbles with volumes $V_i$ and $V_{i+1}$, velocities~$\boldsymbol{u}_i$ and $\boldsymbol{u}_{i+1}$, and positions~$\boldsymbol{x}_i$ and $\boldsymbol{x}_{i+1}$ satisfy the equations
\begin{figure}[H]
	\centering
	\includegraphics[width=\linewidth]{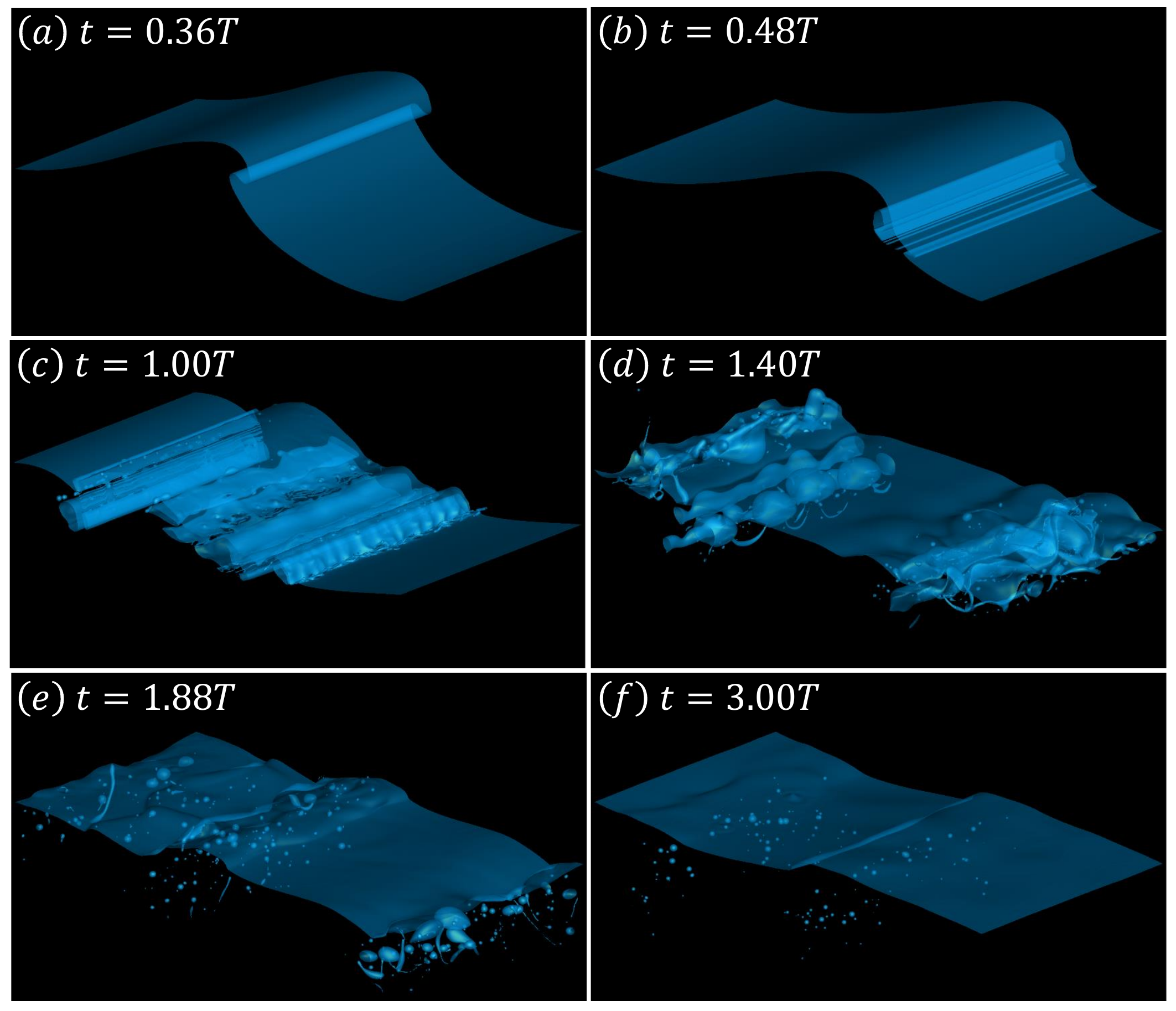}
	\caption{\label{fig:wavesurface55} Visualization of a wave simulation with $S=0.55$ at different times. See the caption of figure~\ref{fig:wavesurface45} for further details.}
\end{figure}
\begin{equation}
    \label{eq:con_vol}
    V_i = V_{i+1}
\end{equation}
and 
\begin{equation}
    \label{eq:con_pos}
    \boldsymbol{x}_i V_i + \frac{1}{2} \left(V_i \boldsymbol{u}_i + V_{i+1} \boldsymbol{u}_{i+1}\right) \Delta t = \boldsymbol{x}_{i+1} V_{i+1},
\end{equation}
where $\Delta t$ is the time interval between the two time instances. Similarly, we can also obtain equations for the binary fragmentation and binary coalescence processes. Based on the constraints imposed on the positions, velocities, and volumes of bubbles between sequential time instances, the ON method defines error functions, namely, pseudodistance error functions. The pseudodistance error function for continuity is defined as
\begin{equation}
    \label{eq:pseudo}
    \Theta_{CN} = \alpha_1 \Delta_p + \alpha_2 \Delta_V + \alpha_3 \Delta_{l},
\end{equation}
with
\begin{equation}
    \label{eq:pse_1}
    \Delta_p^2 = \left\|\boldsymbol{x}_i + \frac{1}{2} \boldsymbol{u}_i \Delta t - \left(\boldsymbol{x}_{i+1} - \frac{1}{2}\boldsymbol{u}_{i+1}\Delta t\right)\right\|^2,
\end{equation}
\begin{figure}[H]
	\centering
	\includegraphics[width=\linewidth]{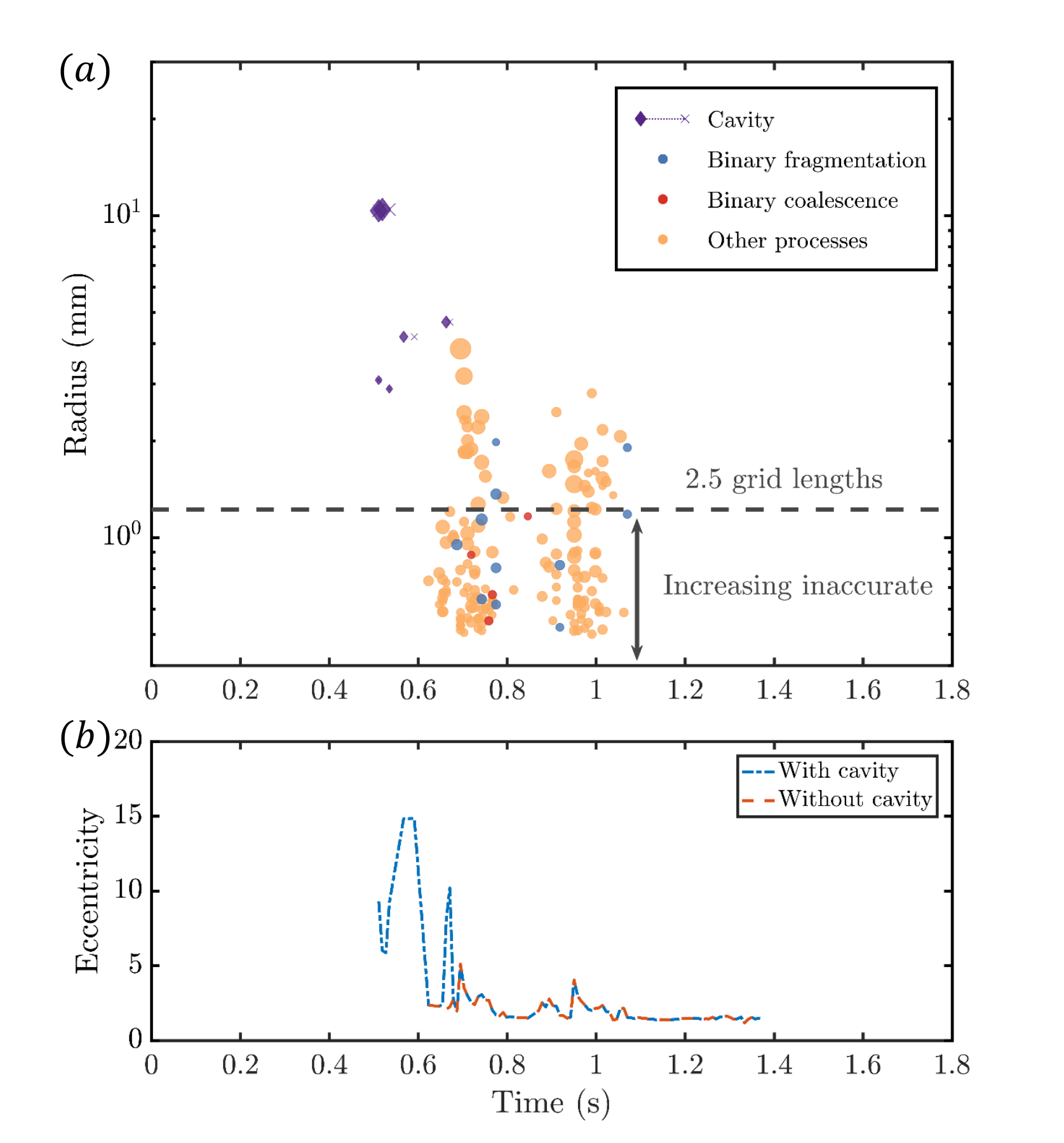}
	\caption{\label{fig:event38} (a):~Bubble events and cylinders in a simulated breaking wave with $S=0.38$. (b):~Average bubble eccentricity as a function of time. In (a), circles denote bubble events, circle colors denote the event type, and size indicates the magnitude of eccentricity. Circles are plotted horizontally with time and vertically with the volume-equivalent radius. Purple diamonds denote cylinders, and their size indicates the cross-sectional area-averaged radius. Diamonds are plotted horizontally by the formation time and vertically by the volume-equivalent radius. The lifetime of a cylinder is indicated by a purple `x' connected to a diamond with a horizontal dotted line. In (b), blue and red lines denote the average bubble eccentricity with and without considering air cavities.}
\end{figure}
\begin{equation}
    \label{eq:pse_2}
    \Delta_V = \frac{\left| V_{i+1} - V_i \right|}{\left(V_i V_{i+1}\right)^{1/3}},
\end{equation}
and
\begin{equation}
    \label{eq:pse_3}
    \Delta_l^2 = \left\|\boldsymbol{x}_i + \frac{1}{2} \boldsymbol{u}_i \Delta t - \left(\boldsymbol{x}_{i+1} - \frac{1}{2}\boldsymbol{u}_{i+1}\Delta t\right)\right\|^2,
\end{equation}
where $\alpha_1$, $\alpha_2$, and $\alpha_3$ are free parameters (good values for these parameters are $0.45$, $0.45$, and $0.1$, respectively). A pseudodistance error function contains three terms: $\Delta_p$, $\Delta_V$, and $\Delta_l$. The first two terms come from equations~\eqref{eq:con_pos} and \eqref{eq:con_vol}. The third term is used to ensure that the events can occur only locally. Similarly, we can define pseudodistance error functions for binary fragmentation and binary coalescence. Here, the formulas for $\Delta_p^2$ and $\Delta_l^2$ are the same for the continuity event. However, this is not the case for fragmentation and coalescence. More details about the formula and their discussion can be found in~Gao el al.~(2021a). All pseudodistance error functions should take minimum values when the bubble events are correctly identified. The ON method makes connections by selecting the minima from a set of pseudodistance error functions. More information about establishing these networks can be found in Gao et al.~(2021a). The robustness of our bubble tracking algorithm has been tested extensively with various simulation cases, and the results show that the accuracies for continuity, binary fragmentation, and binary coalescence are 99.5\%, 90\%, and 95\%, respectively.

\begin{figure}[H]
	\centering
	\includegraphics[width=\linewidth]{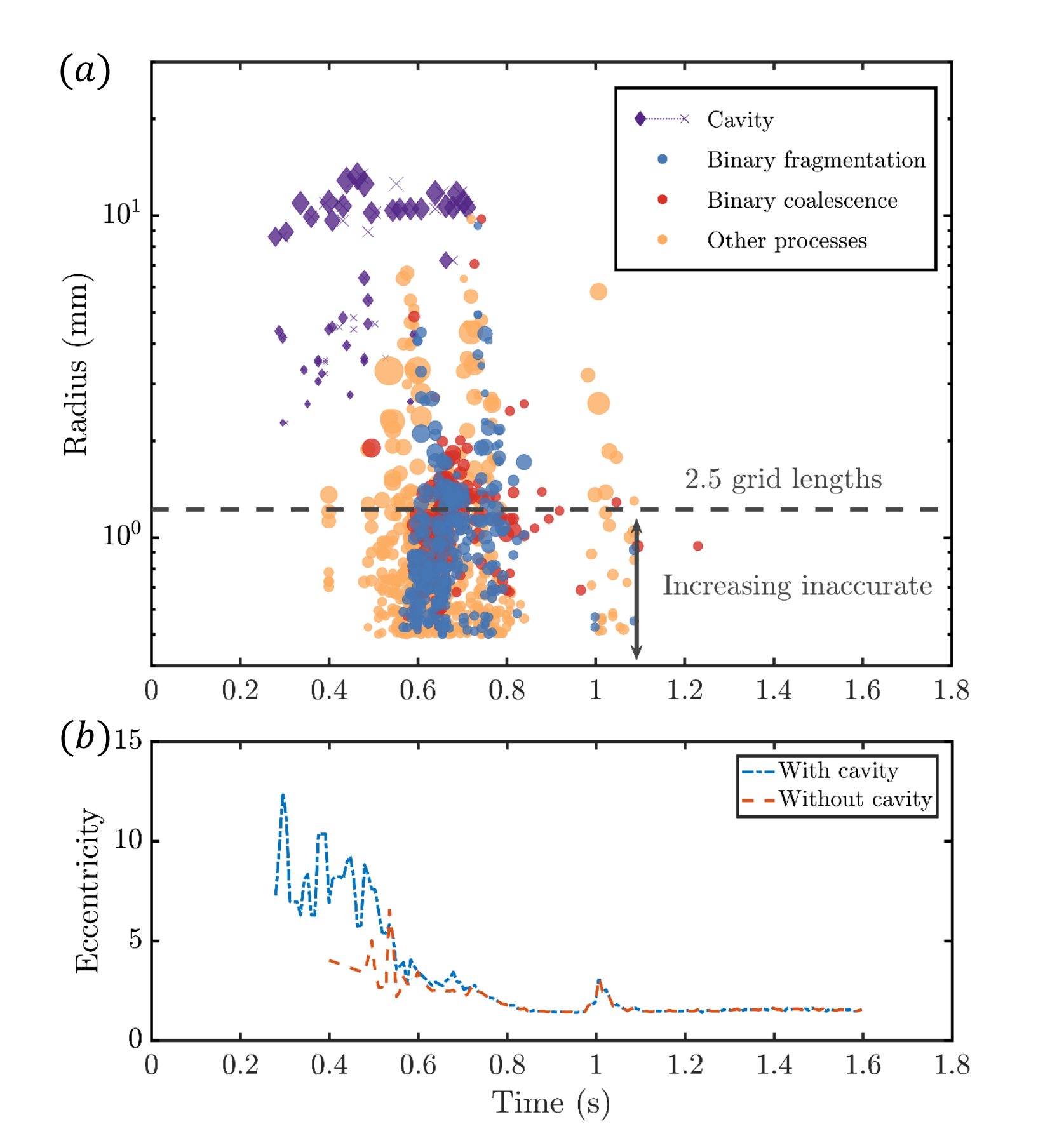}
	\caption{\label{fig:event45} (a):~Bubble events and cylinders in a simulated breaking wave with $S=0.45$. (b):~Average bubble eccentricity as a function of time. See the more detailed description in the caption of figure~\ref{fig:event38} for the meanings of the symbol colors and sizes.}
\end{figure}

\subsection{Sound generated by bubbles}
The underwater noise created within a breaking wave is calculated based on the wave noise model proposed by Deane~(1997) and Deane \& Stokes~(2010). The wave noise model assumes that newly formed bubbles created throughout the acoustically active period of wave breaking are the dominant source of sound and therefore relates the production of sound with these newly formed bubbles, which can be identified using the ON method. In addition, bubble damping effects on underwater noise are considered in this work using the e-folding length and correction factors. More information on the calculation of the sound generated by bubbles can be found in Gao et al.,~(2021b).

\section{Results}
\label{SECresults}
\subsection{Bubble entrainment procedure}
Three cases, viz., $S=0.38$, $S=0.45$, and $S=0.55$, representing types of waves from a spilling breaker to a plunging breaker, are examined here. Visualizations of the breaking wave surface and bubble entrainment process for each of the three cases are plotted in figures~\ref{fig:wavesurface38}, \ref{fig:wavesurface45}, and \ref{fig:wavesurface55}.

The wave breaking and bubble entrainment processes are similar among the three wave slopes simulated herein. Initially, a jet forms on the wave crest. Then, the overturning wave crest strikes the wave surface, producing multiple cavities and splash jets. Filaments and cavities eventually break up into bubbles due to the instability of air cylinders, and larger bubbles fragment into smaller bubbles via a fragmentation cascade. Finally, larger bubbles rise toward the wave surface faster than the smaller bubbles, leaving the latter suspended underwater for a relatively long time. With increasing wave steepness, the wave breaking process becomes more energetic, leading to the creation of more bubbles. In addition, bubbles are entrained at shallower depths when $S=0.38$ than when $S=0.55$. As shown in figures~\ref{fig:wavesurface38}, \ref{fig:wavesurface45}, and \ref{fig:wavesurface55}, two bubble creation mechanisms are observed: air filament/cavity breakup and turbulence fragmentation cascade. The air filament/cavity breakup happens at the early stage of the breaking waves when the turbulence is weak and produces bubbles of various sizes, which is followed by the process that relatively large bubbles fragment into smaller ones due to turbulence fluctuations. Comparing the simulation results for different wave steepness cases, the turbulence intensity increases as the wave steepness increases. This might imply that the fragmentation cascade process is more important for the high steepness case, whereas the air filament/cavity breakup seems more important for the low steepness case. For example, for the low wave steepness case ($S=0.38$), from figure~\ref{fig:wavesurface38}, most of the bubbles are created by the air filament/cavity breakup. It is very difficult to develop an algorithm that can robustly distinguish the bubbles created by different mechanisms. Therefore, in this work, we only provide qualitative descriptions of the two bubble creation mechanisms. Detailed quantitative and accurate analysis will be in the future study.

\begin{figure}[H]
	\centering
	\includegraphics[width=\linewidth]{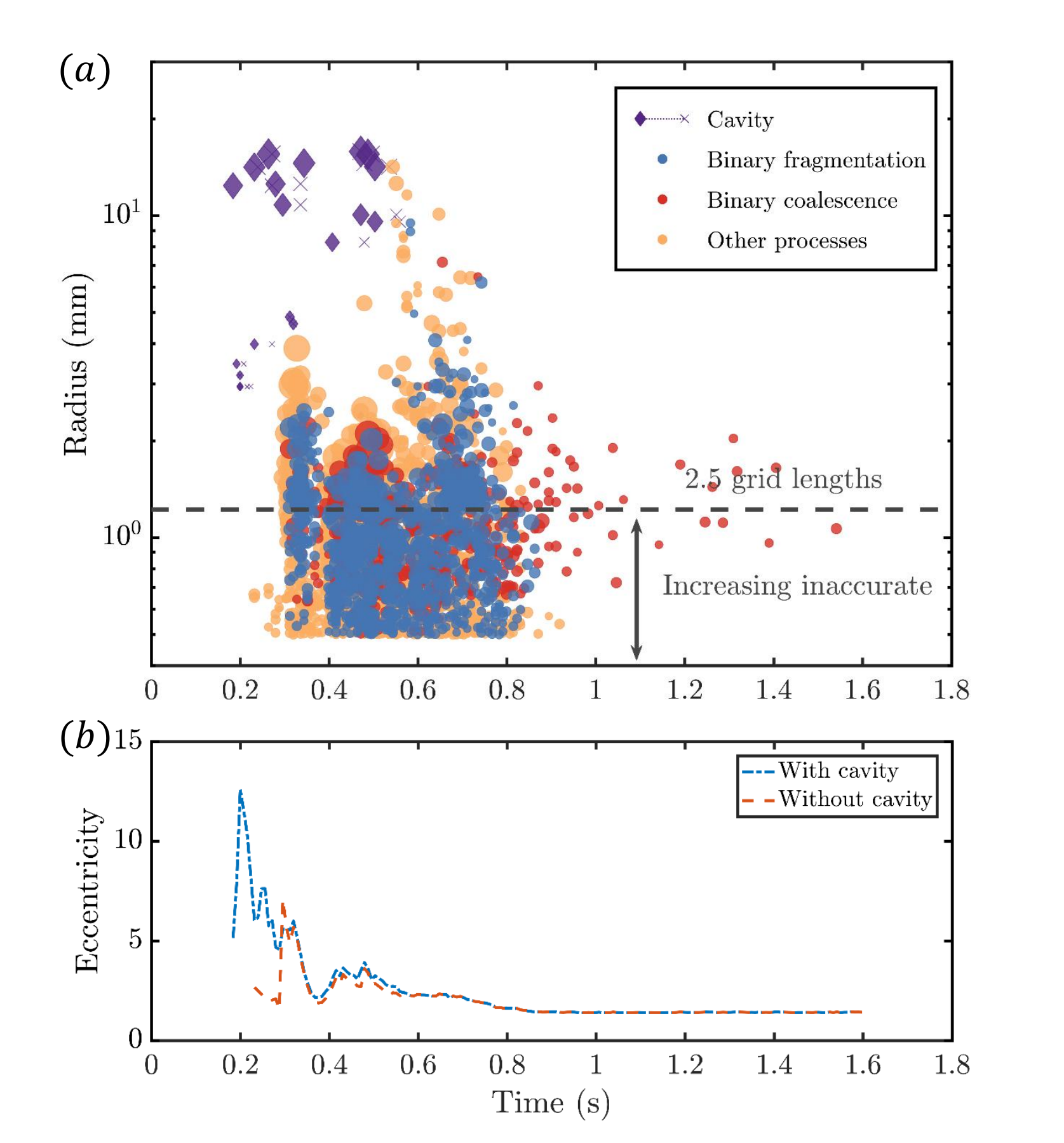}
	\caption{\label{fig:event55} (a):~Bubble events and cylinders in a simulated breaking wave with $S=0.55$. (b):~Average bubble eccentricity as a function of time. See the more detailed description in the caption of figure~\ref{fig:event38} for the meanings of the symbol colors and sizes.}
\end{figure}

Bubble creation events can be robustly and accurately detected by the ON method. The bubble creation events detected by the ON method for the cases $S=0.38$, $S=0.45$, and $S=0.55$ are plotted in figures~\ref{fig:event38}(a), \ref{fig:event45}(a), and \ref{fig:event55}(a), respectively. Three categories for bubble behaviors, namely, binary fragmentation, binary coalescence, and other processes, and two categories for cavity behaviors, namely, creation and extinction, are considered here. A cavity is defined as an air cylinder that spans across the whole simulation domain in the spanwise direction. A cavity is represented by a purple diamond connected to an `x' with a horizontal dotted line, where the diamond and `x' denote creation and extinction events, respectively, and the sizes of the diamond and `x' represent the cross-sectional area-averaged radius of the cylinder. Blue, red, and yellow circles represent binary fragmentation events, binary coalescence events, and other processes, respectively. These other processes, including multiple fragmentation, multiple coalescence, and direct entrainment, are the events that the ON method cannot identify. The circle size denotes the bubble eccentricity, which is defined as
\begin{equation}
    \label{eq:eccentricity}
    e = \frac{d_{max}}{r_v},
\end{equation}
where $d_{max}$ is the maximum distance from a point in the bubble to the center of the volume-equivalent spherical bubble and $r_v$ is the volume-equivalent radius of the bubble. The eccentricity describes the degree to which the shape of the bubble is irregular and takes a value of $1$ for a perfectly spherical bubble. The dashed lines in figures~\ref{fig:event38}(a), \ref{fig:event45}(a), and \ref{fig:event55}(a) show a scale equal to $2.5$ grid lengths. Note that for bubbles smaller than this scale, the physics of the surface tension become inaccurate in the simulation. Here we still plotted the data points for bubbles smaller than this scale, however, caution should be taken and higher resolution simulations are needed for dealing with these bubbles.

\begin{figure}[H]
	\centering
	\includegraphics[width=\linewidth]{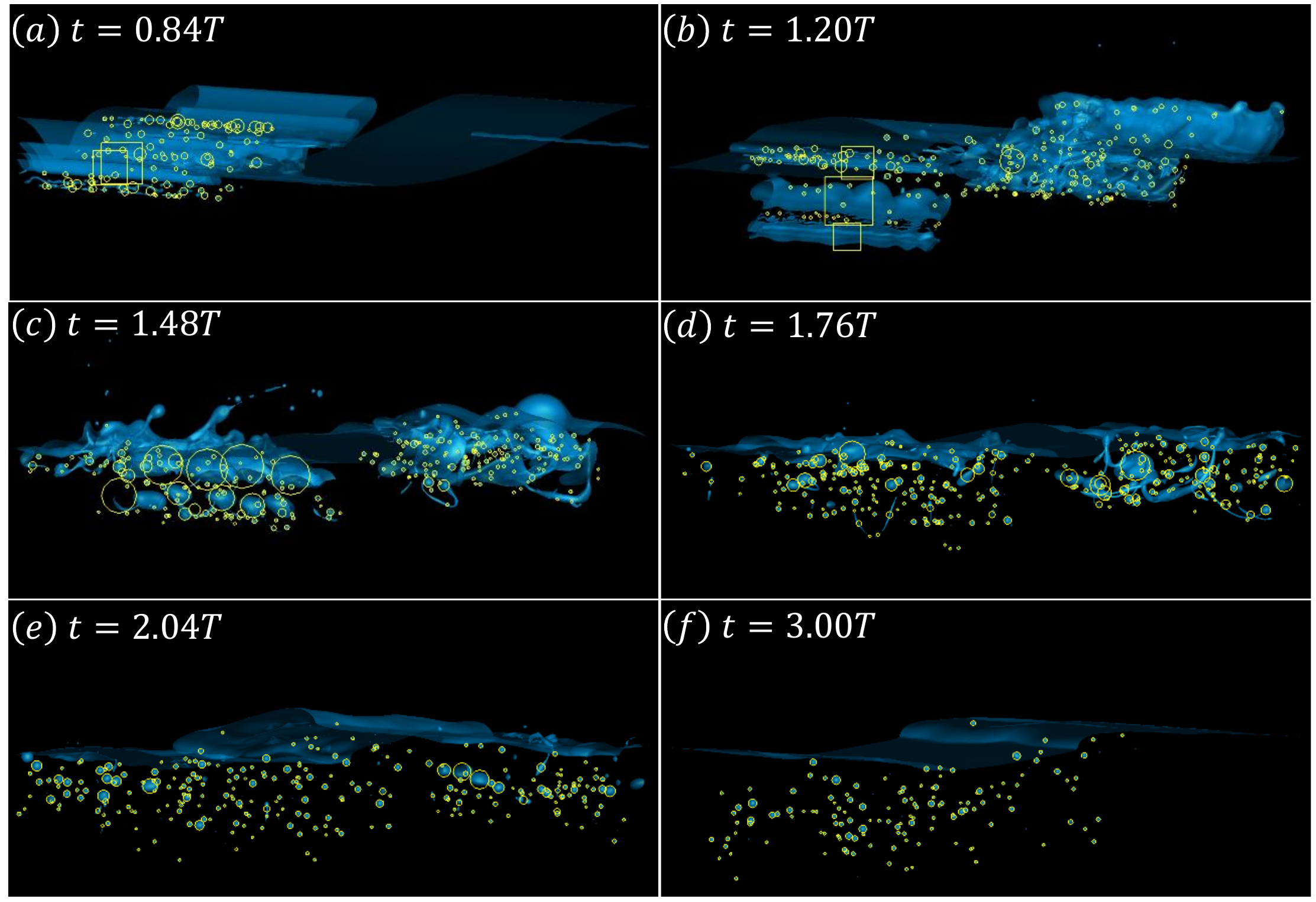}
	\caption{\label{fig:wavesurfaceid} Identified bubbles and cavities in a breaking wave~($S=0.55$). The yellow circles and squares denote identified bubbles and air cylinders, respectively. (a)--(c):~Air cylinders break up into multiple bubbles. (d):~Larger bubbles fragment into smaller bubbles, which is referred to as a bubble fragmentation cascade. (e)--(f):~Degassing process.}
\end{figure}
Figure~\ref{fig:event38} demonstrates the entrainment of a few air cylinders, followed by two clusters of bubbles, which are created by two groups of entrained air cylinders. Some air cylinders are not identified as air cavities because they do not span across the whole domain in the spanwise direction. Most bubbles are created by other processes because they are associated mainly with cylinder instability and thus cannot be identified by the ON method. Compared to those when $S=0.38$, the breaking waves are more energetic and produce more bubbles when $S=0.45$ and $S=0.55$. A number of cylinders are entrained at the onset of wave breaking and subsequently disappear when many bubbles are created by binary fragmentation, binary coalescence, and other processes. Initially, bubbles are created with irregular shapes, but they become more spherical over time. Smaller bubbles tend to have small eccentricities because surface tension is more important on small scales. Figure~\ref{fig:event55} clearly shows that, as expected, the largest bubble size decreases over time, which is the result of a fragmentation cascade.

\begin{figure}[H]
	\centering
	\includegraphics[width=\linewidth]{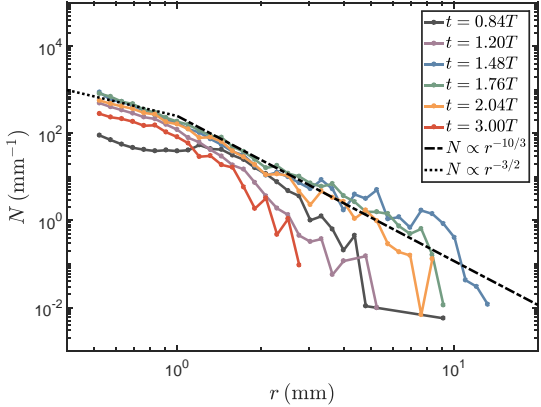}
	\caption{\label{fig:ensemble_size} Ensemble-averaged bubble size spectrum. Color denotes time. The dotted line and dashed--dotted line correspond to the power-law scalings of $-3/2$ and $-10/3$ representing the theoretical predictions for sub-Hinze-scale bubbles and super-Hinze-scale bubbles, respectively.}
\end{figure}

The average bubble eccentricities as a function of time for the cases $S=0.38$, $S=0.45$, and $S=0.55$ are plotted in figures~\ref{fig:event38}(b),~\ref{fig:event45}(b), and \ref{fig:event55}(b), respectively. The average eccentricity is defined as
\begin{equation}
    \label{eq:ecc_time}
    \bar{e}=\frac{\sum{r_ve}}{\sum{r_v}},
\end{equation}
where $\sum$ denotes the summation over all the bubbles in the simulation domain. In figures~\ref{fig:event38}(b),~\ref{fig:event45}(b), and \ref{fig:event55}(b), the blue and red lines represent the average eccentricities with and without cavities, respectively. A comparison between panels (a) and (b) in figures~\ref{fig:event38},~\ref{fig:event45}, and \ref{fig:event55} generally suggests that the eccentricity peaks correspond to cylinder breakup and bubble creation events. Then, the average eccentricity decreases after the peak, indicating that the bubbles are entrained with irregular shapes and become more spherical over time. This result reveals that cylinder instability might be critical for the transition from the two-dimensional wave field to the three-dimensional turbulent field of a breaking wave.

\subsection{Bubble size spectrum}
The bubble size spectrum is an important physical quantity for quantitatively describing bubble entrainment. Garrett et al.~(2000) stated that breaking waves initially entrain large bubbles, which subsequently break up into smaller bubbles. This process is referred to as a bubble fragmentation cascade. Considering that bubble entrainment is associated with the average rates of the supply of air and energy dissipation, Garrett et al.~(2000) conducted a dimensional analysis and proposed that the bubble size spectrum shows a $-10/3$ power-law scaling for bubbles with radii greater than the Hinze scale, which is the scale at which the bubble surface tension balances the turbulent fluctuation.
\begin{figure}[H]
	\centering
	\includegraphics[width=\linewidth]{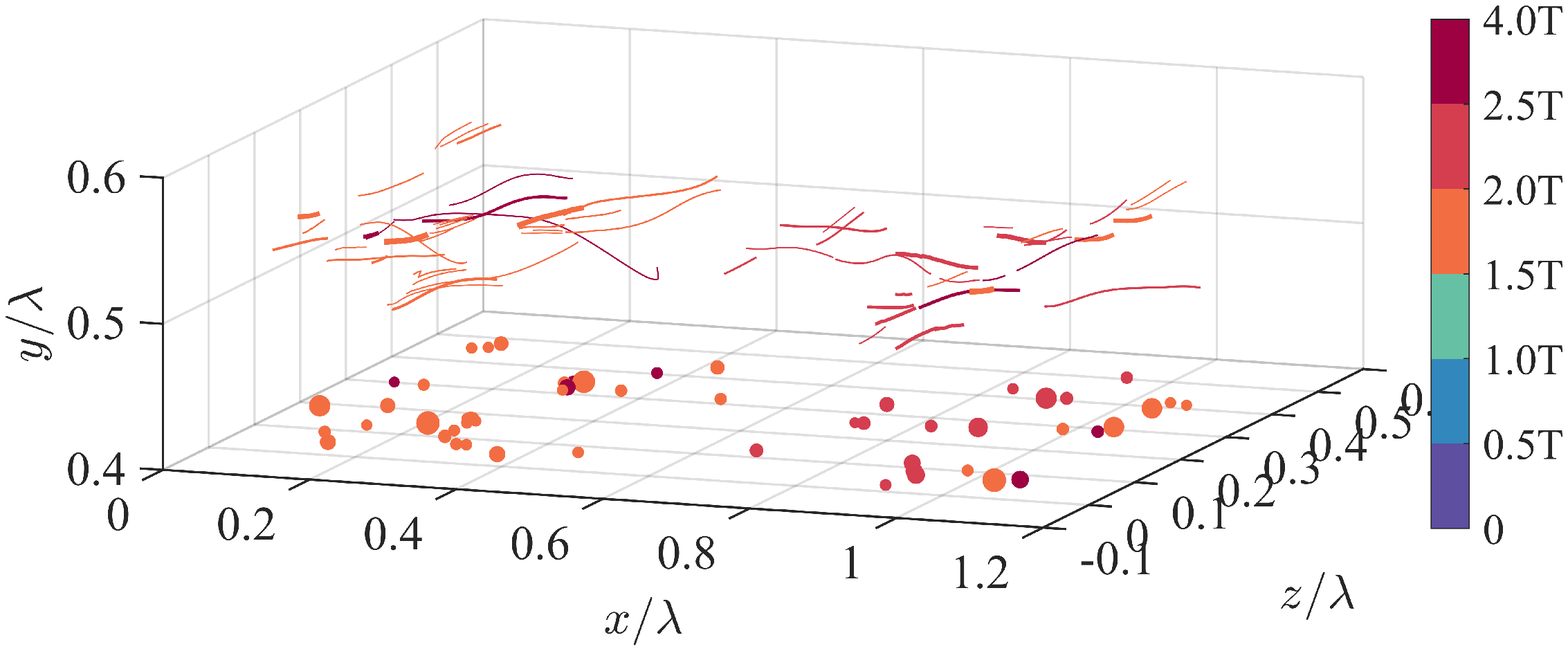}
	\caption{\label{fig:traj_3d_38} Oblique view of the bubble trajectories for the case~$S=0.38$. Only bubble trajectories with bubble radii greater than $0.5\,\mathrm{mm}$ and residence times greater than $0.04T$ are plotted here. Line color and width denote the bubble creation time and bubble size, respectively. The scatters in the bottom panel represent the bubble bursting locations, and the scatter size is proportional to the bubble size.}
\end{figure}
\begin{figure}[H]
	\centering
	\includegraphics[width=\linewidth]{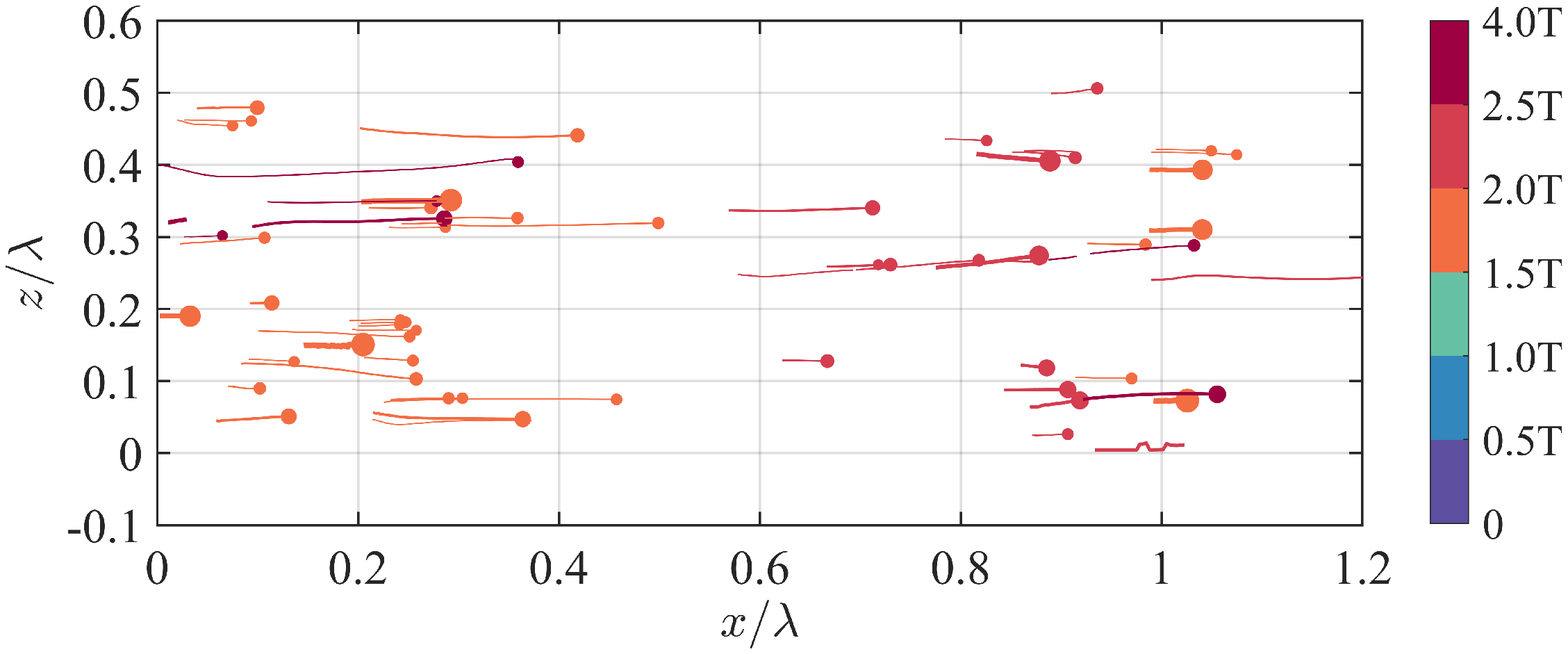}
	\caption{\label{fig:traj_topview_38} Top view of the bubble trajectories for the case~$S=0.38$. For more detailed descriptions of the meanings of the line color, line width, scatter location, and scatter size, see the caption of figure~\ref{fig:traj_3d_38}.}
\end{figure}
Deane \& Stokes~(2002) conducted experiments and verified the $-10/3$ power-law scaling for super-Hinze-scale bubbles~(bubbles whose radii exceed the Hinze scale) at the end of the active phase of air entrainment. Moreover, they found that sub-Hinze-scale bubbles~(bubbles whose radii are smaller than the Hinze scale) follow a $-3/2$ power-law scaling. However, the detailed mechanism of sub-Hinze-scale bubble entrainment remains unclear. Figure~\ref{fig:wavesurfaceid} shows the bubble entrainment process, and figure~\ref{fig:ensemble_size} shows the corresponding ensemble-averaged bubble size spectrum. The definition of the bubbles used in calculating the size spectrum is consistent with the circles in figure~\ref{fig:event55}(a). The fragmentation model proposed in Garrett et al.~(2000), which leads to a -10/3 power-law scaling for the bubble size spectrum, embodies an average air supply rate or a source volume due to the entrainment process during wave breaking. The recent work of Gaylo et al.~(2021) defines an entrainment size distribution of bubble sources which reproduces the same -10/3 scaling for the equilibrium bulk bubble size spectra (for super-Hinze scale bubbles) via fragmentation cascading, in the weak volume injection regime. In our current study, using the CCL algorithm mentioned in the preceding section, we are able to identify the spanwise air cylinders or cavities entrained by the breaking wave during initial stages, and treat them to be the primary source volumes which undergo breakup to generate bubbles. However, demarcating and estimating the bubble injection size distribution due to air entrainment in a breaking wave, and studying its subsequent evolution into the bulk bubble size spectrum by employing the bubble identification and tracking subroutines, is a future topic of interest.
\begin{figure}[H]
	\centering
	\includegraphics[width=\linewidth]{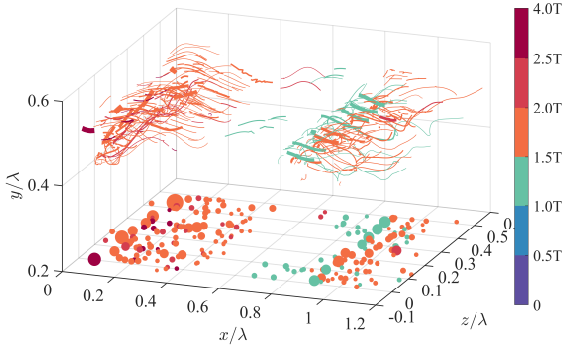}
	\caption{\label{fig:traj_3d_45} Oblique view of the bubble trajectories for the case $S=0.45$. For more detailed descriptions of the meanings of the line color, line width, scatter location, and scatter size, see the caption of figure~\ref{fig:traj_3d_38}.}
\end{figure}
In figure~\ref{fig:wavesurfaceid}, the yellow circles and rectangles correspond to bubbles and air cylinders, respectively. Air cavities and filaments are entrained in the early stage of wave breaking (see figures~\ref{fig:wavesurfaceid}(a) and 7(b)). At $t=0.84T$ and $t=1.20T$, the slopes of the bubble size spectrum~(black line and purple line, respectively, in figure~\ref{fig:ensemble_size}) do not follow the $-10/3$ power-law scaling for super-Hinze-scale bubbles because some of the air filaments have broken up into bubbles of different scales due to the abovementioned cylinder instability~(Gao et al., 2021b), which is completely different from a fragmentation cascade. At $t=1.48T$, $t=1.76T$, and $t=2.04T$, air cavities break up and create large bubbles, and some of these large bubbles fragment into smaller bubbles; this phenomenon represents the fragmentation cascade process described above. In addition, because large bubbles have higher ascension speeds than small bubbles, some of these large bubbles rise up and burst on the wave surface. The turbulence saturation hypothesis~(Deane et al. 2016) implies that the bubble Hinze scale shows less dependence on the wave scales and turbulence energy dissipation. Therefore, we expect that the Hinze scale in this study is about $1\,\mathrm{mm}$. We can clearly see that the bubble size spectrum for a fragmentation cascade shows a $-10/3$ power-law scaling for bubbles greater than the Hinze scale, which is consistent with the conclusions in the previous literature. The bubble size spectrum at the Hinze scale seems showing a transition power-law scaling. This point has been verified by the grid convergence study in the authors' previous work and will not be repeated here. More information can be found in figure~8 in Gao et al.~(2021b).
\begin{figure}[H]
	\centering
	\includegraphics[width=\linewidth]{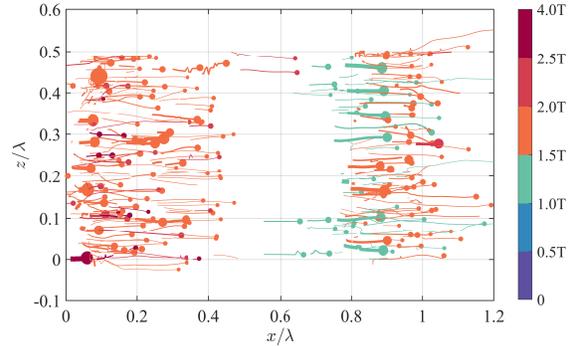}
	\caption{\label{fig:traj_topview_45} Top view of the bubble trajectories for the case $S=0.45$. For more detailed descriptions of the meanings of the line color, line width, scatter location, and scatter size, see the caption of figure~\ref{fig:traj_3d_38}.}
\end{figure}
At $t=3.00T$, the slope of the size spectrum is steeper than $-10/3$~(red line in figure~\ref{fig:ensemble_size}) because the large bubbles have disappeared, leaving only small bubbles in the water that persist for a relatively long time. The temporal evolution of the bubble size spectrum shows that the bubble creation mechanism due to cylinder instability~(Gao et al.,~2021) has significant impacts on the bubble size spectrum through the following aspects. First, the bubble size spectrum at the early wave breaking stage ($t=0.84T$ and $t=1.20T$) does not follow the $-10/3$ power-law scaling because the bubble creation mechanism is not a turbulence-driven fragmentation cascade. We discover that some air cylinders break up into small bubbles, which might indicate that cylinder instability is critical during this stage. Second, as shown in figure~\ref{fig:ensemble_size}, the two peaks near $r=5\,\mathrm{mm}$ and $r=8\,\mathrm{mm}$ in the bubble size spectrum at $t=1.48T$~(blue line) are created by the breakup of an air cavity (see figure~\ref{fig:wavesurfaceid}(c)). Finally, the relatively large bubbles created by the breakup of an air cavity are further broken up through a fragmentation cascade, indicating that these large bubbles 
are important in the bubble size spectrum at the end of the acoustically active phase.
\begin{figure}[H]
	\centering
	\includegraphics[width=\linewidth]{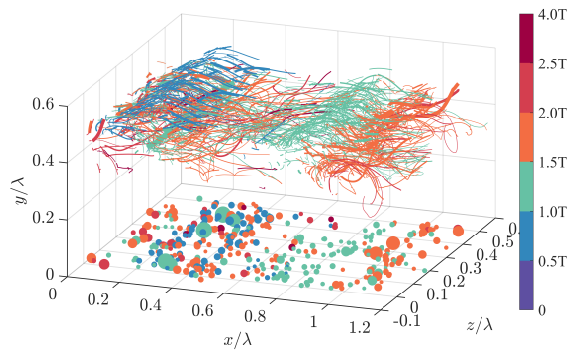}
	\caption{\label{fig:traj_3d_55} Oblique view of the bubble trajectories for the case $S=0.55$. For more detailed descriptions of the meanings of the line color, line width, scatter location, and scatter size, see the caption of figure~\ref{fig:traj_3d_38}.}
\end{figure}
\begin{figure}[H]
	\centering
	\includegraphics[width=\linewidth]{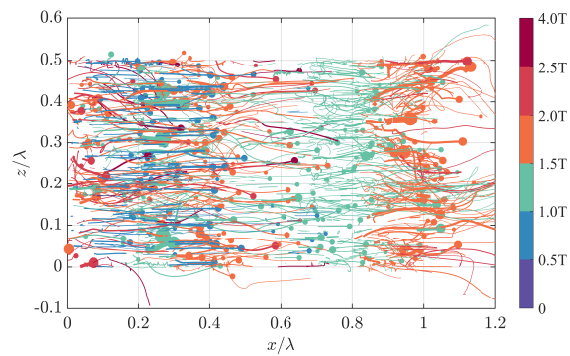}
	\caption{\label{fig:traj_topview_55} Top view of the bubble trajectories for the case $S=0.55$. For more detailed descriptions of the meanings of the line color, line width, scatter location, and scatter size, see the caption of figure~\ref{fig:traj_3d_38}.}
\end{figure}
Bubble trajectories play important roles in many ocean--atmosphere processes, such as air--sea gas transfer and sea spray generation. Compared with nonbreaking waves, breaking waves and bubbles increase the air--water interface area, resulting in an enhanced air--sea gas transfer rate. In addition, bubbles burst on the wave surface at the end of some bubble trajectories and produce film drops and jet drops. For these and other reasons, detailed investigations of bubble trajectories are necessary for ongoing research. Nevertheless, there is no literature reporting the trajectories of bubbles in breaking waves using a direct numerical simulation because of the difficulties in tracking bubbles and detecting their evolving behavior over time. However, the recent progress afforded by the bubble tracking algorithm developed by~Gao et al.~(2021a), namely, the ON method, enables us to accurately and robustly detect and track bubble events. To identify bubble trajectories, the connections among continuous bubbles are established over time.
\begin{figure}[H]
	\centering
	\includegraphics[width=\linewidth]{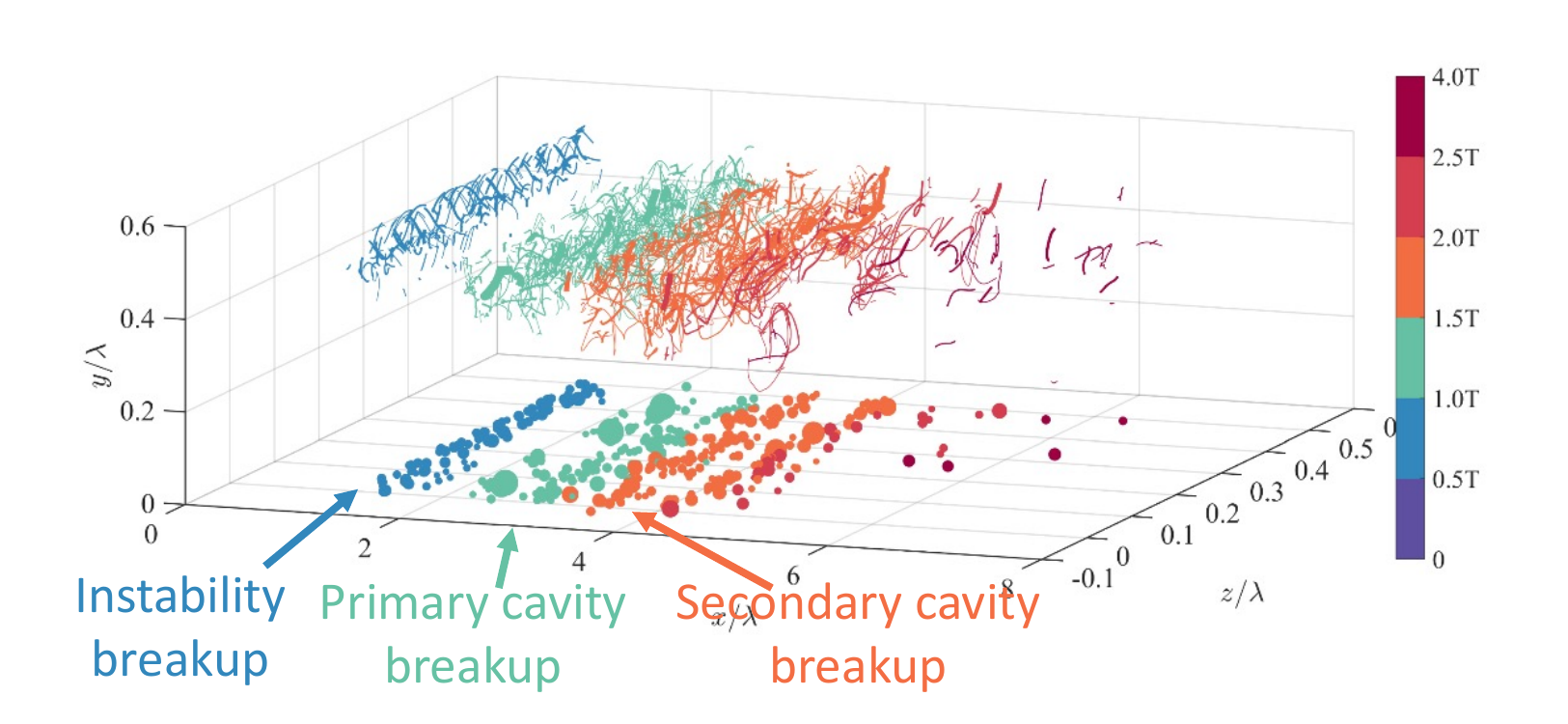}
	\caption{\label{fig:trajectory55} Oblique view of the bubble trajectories for the case $S=0.55$ with bubble locations shifted according to their creation time. For more detailed descriptions of the meanings of the line color, line width, scatter location, and scatter size, see the caption of figure~\ref{fig:traj_3d_38}.}
\end{figure}
The oblique and top views of the bubble trajectories for the cases $S=0.38$, $S=0.45$, and $S=0.55$ are plotted in figures~\ref{fig:traj_3d_38}--\ref{fig:traj_topview_55}. Only the trajectories of bubbles whose radii exceed $0.5\,\mathrm{mm}$ and whose residence times are greater than or equal to $0.04T$ ($T$ is the wave period) are plotted here. The color signifies the bubble creation time. The line width and scatter size denote the bubble size, and the scatter location reflects the bubble bursting location. As shown in figures~\ref{fig:traj_3d_38} and \ref{fig:traj_topview_38}, only a few bubbles are created in the water, and the bubble trajectories are relatively short and straight. For the cases $S=0.45$ and $S=0.55$, more bubbles are created than in the case $S=0.38$, and the bubble trajectories are more chaotic and arcuate with some trajectories exhibiting rotation. The bubble trajectories have three unique features: parallel curves, rotational structures, and tails pointing upward and forward. For example, we observe the parallel green lines at $x/\lambda \approx 0.9$ in figures~\ref{fig:traj_3d_45} and \ref{fig:traj_topview_45} and the parallel blue lines at $x/\lambda \approx 0.3$ in figures~\ref{fig:traj_3d_55} and \ref{fig:traj_topview_55}. The parallel structures show that the bubbles are created at almost the same time with similar sizes and are distributed uniformly in the spanwise direction, supporting the hypothesis of bubble production by the cylinder instability mechanism described in~Gao et al.~(2021b). In contrast, the rotational orange curves at $x/\lambda\approx 1$ in figure~\ref{fig:traj_3d_55} are the result of the interaction between the bubble trajectories and underwater flow vortex structures. In addition, owing to buoyancy, some bubbles rise up while being transported; as a result, the trajectory tails of some large bubbles point upward and forward. These bubbles are generally the large ones that the vortex structures are not able to capture. {If we shift the bubble trajectories location according to their creation time, figure~\ref{fig:traj_3d_55} can be replotted as figure~\ref{fig:trajectory55}, where the bubble and air cylinder events can be clearly observed.}
\begin{figure}[H]
	\centering
	\includegraphics[width=\linewidth]{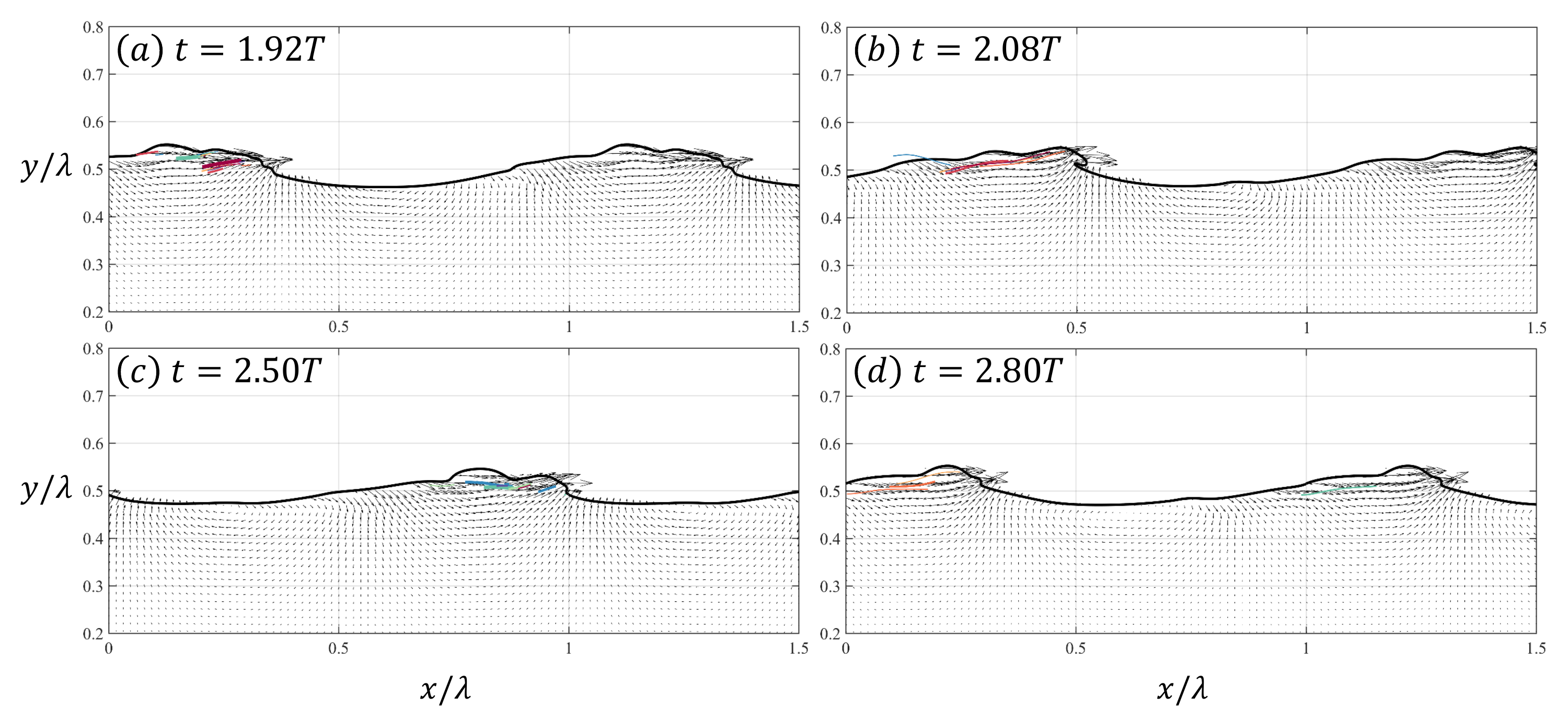}
	\caption{\label{fig:traj_vortex_38} Interaction between the bubble trajectories and flow field for the case~$S=0.38$. The velocity field is plotted as the vectors with the spanwise-averaged components $u$ and $v$. The averaged wave surface is plotted as the black line, which is the zeroth isosurface of the spanwise-averaged LS function. The colored lines are the bubble trajectories. For more detailed descriptions of the meanings of the color and width of the trajectories, see the caption of figure~\ref{fig:traj_3d_38}.}
\end{figure}
\begin{figure}[H]
	\centering
	\includegraphics[width=\linewidth]{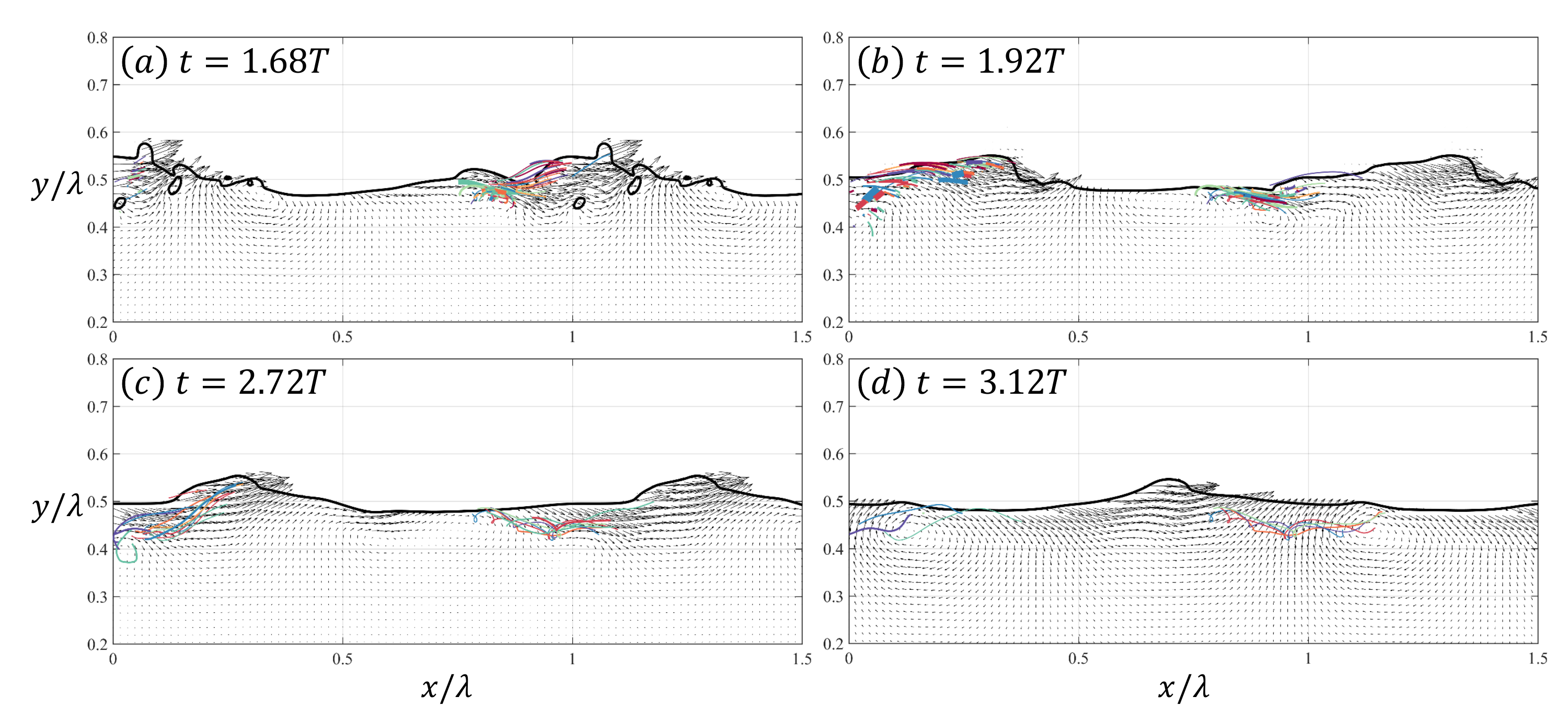}
	\caption{\label{fig:traj_vortex_45} Interaction between the bubble trajectories and flow field for the case~$S=0.45$. For more details about the meanings of the vectors and lines, see the caption of figure~\ref{fig:traj_vortex_38}.}
\end{figure}
To check the interactions between the bubble trajectories and flow field, the spanwise-averaged velocities and the bubble trajectories are plotted in figures~\ref{fig:traj_vortex_38}, \ref{fig:traj_vortex_45}, and \ref{fig:traj_vortex_55} for the cases~$S=0.38$, $S=0.45$, and $S=0.55$, respectively. For the small wave steepness case~($S=0.38$), as shown in figure~\ref{fig:traj_vortex_38}, bubbles are entrained at shallow depths and quickly merge with the wave surface, resulting in relatively short, straight trajectories. To illustrate the vortex structures in breaking waves, a straightforward way is to use the vorticity contours. However, to show the colored trajectories and their interaction with large vortex structure clearly, here, we use the velocity vectors to show the large vortex structures. As shown in figures~\ref{fig:traj_vortex_45} and \ref{fig:traj_vortex_55}, for the cases~$S=0.45$ and $S=0.55$, the rotational vortex structures clearly affect the transport of bubbles and produce rotational bubble trajectories.  Nevertheless, in this work, we only qualitatively describe the interactions between the bubble trajectories and the flow field, and more detailed research is needed in the future.
\begin{figure}[H]
	\centering
	\includegraphics[width=\linewidth]{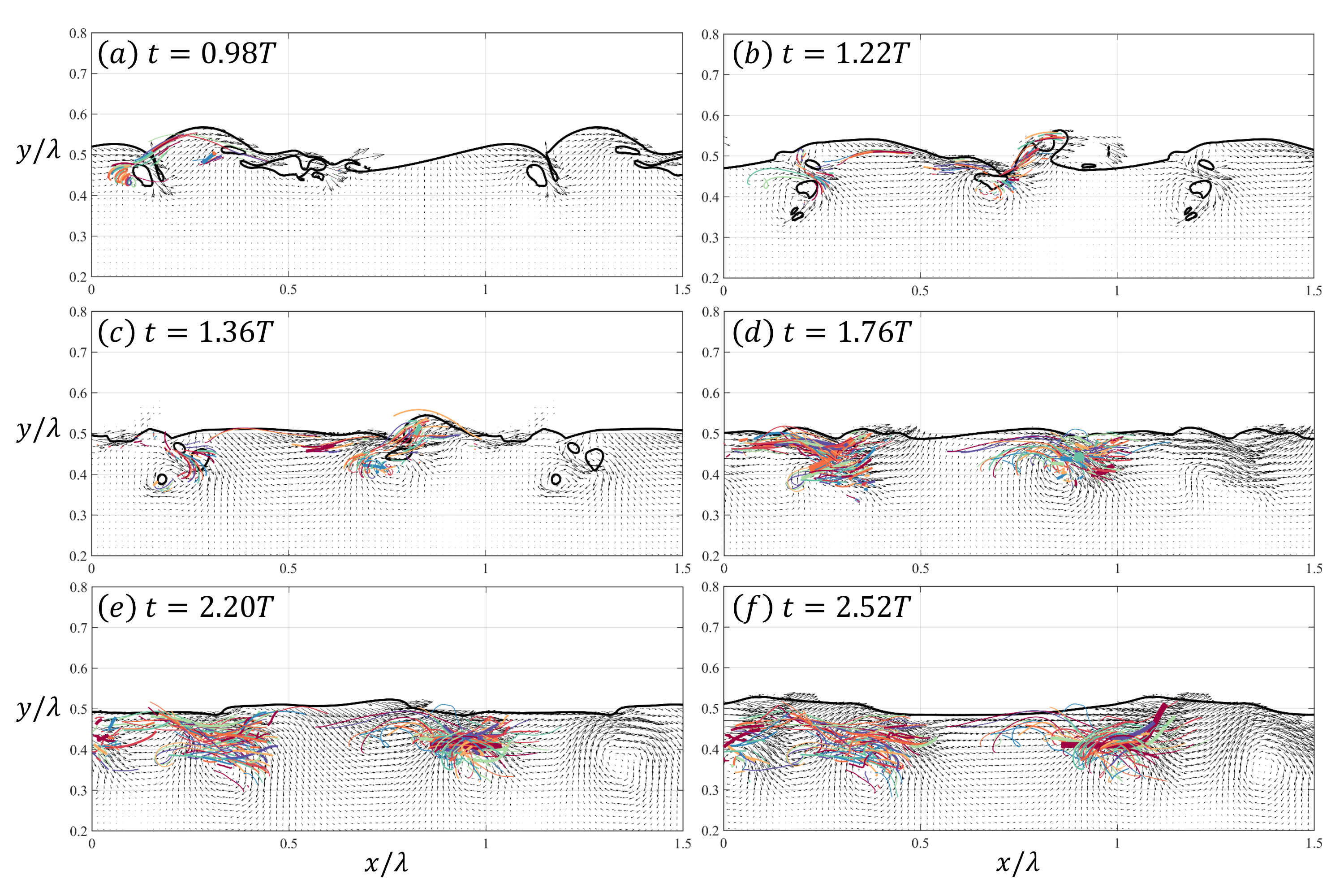}
	\caption{\label{fig:traj_vortex_55} Interaction between the bubble trajectories and flow field for the case~$S=0.55$. For more details about the meanings of the vectors and lines, see the caption of figure~\ref{fig:traj_vortex_38}.}
\end{figure}
\subsection{Wave noise spectrogram}
\begin{figure}[H]
	\centering
	\includegraphics[width=\linewidth]{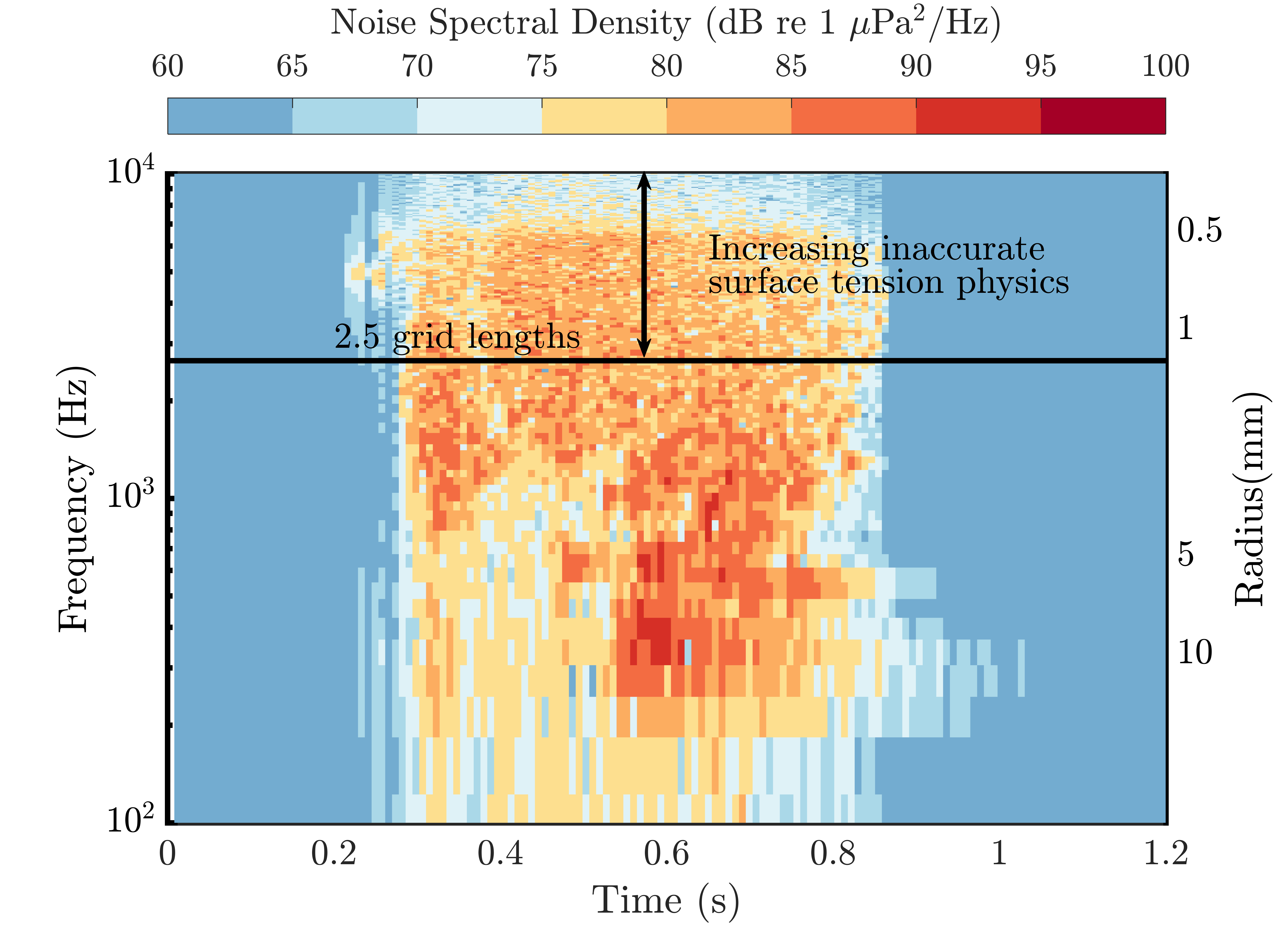}
	\caption{\label{fig:spectrogram} Wave noise spectrogram for the case $S=0.55$.}
\end{figure}

The bubble creation events for the case $S=0.55$ are shown in figure~\ref{fig:event55}. Bubble acoustics are strongly associated with bubble creation events. Based on a dataset of bubble creation events and using the wave noise calculation method of Gao et al.~(2021c), we calculate and investigate the wave noise spectrogram for the case $S=0.55$. The spectrogram is plotted in figure~\ref{fig:spectrogram}. The wave noise energy is concentrated in the time domain during $0.3\,\mathrm{s}$--$0.8\,\mathrm{s}$, which is the acoustically active phase, and in the frequency domain within $330\,\mathrm{Hz}$--$6600\,\mathrm{Hz}$, which corresponds to bubble radii ranging from $10\,\mathrm{mm}$ to $0.5\,\mathrm{mm}$. At approximately $0.6\,\mathrm{s}$, the high intensity of low-frequency wave noise indicates the large bubbles produced by air cavities. In figure~\ref{fig:spectrogram}, we plot a solid line denoting the $2.5$ grid length and highlight that the wave frequencies above this line are not accurate due to the increasingly inaccurate surface tension calculation attributable to the grid resolution limitation. More results and the comparison between simulation results with experimental datasets on wave noise power spectral density can be found in~Gao et al.~(2021c).
\section{Discussions}
\label{SECdiscussion}
In this work, we investigated bubble creation mechanisms and bubble acoustic characteristics for breaking waves. Specifically, a new model was developed to relate bubble behaviors to underwater sound generation, and new physical insight was found that drives bubble production within breaking wave crests in this work. Our bubble acoustic model and the new insights on bubble creation mechanisms can also be applied to other bubble-associated physical phenomena, for example, breaking wave and bubble generation induced by vortex/interface interaction or wave--body interaction. Hendrickson et al. (2022) developed an air entrainment volume for the quasi-two-dimensional interactions of rising surface-parallel vorticity with an air--water interface and applied their model to the quasi-steady wave breaking behind a fully submerged horizontal circular cylinder. Some air cylinders (air filament or cavity) similar to what we observed in the breaking waves, can also be found in their simulations (See figures~3 and 8 in Hendrickson et al., 2022). This implies that our theory for predicting the bubble creation induced by air cylinder breakup is applicable to their simulations. It will be also interesting to check bubble acoustics of the wave--body interaction using our bubble acoustic model in future work.

The ON method can only detect continuity, binary fragmentation, and binary coalescence. Therefore, in this work, we have only shown the results regarding that can be robustly obtained by the ON method, for example, the bubble trajectories and the bubble creation rate (wave spectrogram). However, bubbles produced by cylinder instability involve multiple fragmentation, where the ON method fails. Thus, we cannot distinguish the bubbles produced by different mechanisms, such as bubble fragmentation cascade and cylinder instability, using the ON method. A more advanced algorithm that can track bubble multiple fragmentation and multiple coalescence is needed to provide more details on the bubble statistics created by cylinder instability quantitatively.

The wave profile in the initial condition is uniform in the spanwise direction. It is interesting to know the bubble creation and bubble statistics if the perturbation is added along the spanwise direction. If the perturbation is small, the most unstable mode of the perturbation grows fastest based on the stability theory and the bubble produced by air cylinder instability should still follow our air cylinder breakup theory. However, if the perturbation is large enough, the final bubble sizes should depend on the initial perturbation heavily. However, we expect that the perturbation might influence the number of bubbles created, but it should not affect the bubble size spectrum power-law scaling slope, because the power-law scaling seems a strong and uniform law across a wide range of wave scales and various types of breaking waves verified both experimentally and numerically. Deike et al.~(2016) performed breaking wave simulations with spanwise perturbation in the initial wave profile. Their results show that the perturbation has little effects on the bubble dynamics and the bubble statistics converge to the power-law scaling. More simulations and studies will be performed to investigate the initial perturbation effects in future work.

\section{Conclusions}
\label{SECconclusion}
In this work, we adopted a novel algorithm to track bubbles and detect their evolutionary behaviors over time during breaking wave simulations and investigated various bubble creation mechanisms and bubble trajectories. The results show that instability along the boundary of an air cylinder is an important mechanism for bubble creation and has significant effects on the initial breaking wave process and bubble size spectrum. Bubble trajectories show three unique patterns in breaking waves. Specifically, we observed trajectories with rotational motion, trajectories pointing forward and upward, and parallel trajectories. Finally, we investigated the acoustics of newly formed bubbles and calculated the noise spectrogram radiated by a breaking wave. In future work, we will conduct more detailed studies on bubble creation mechanisms, bubble trajectories interacting with breaking wave flow fields, and wave noise.

\section{Acknowledgements}
\label{SECacknowledgements}
The UMN team's research was previously supported partially by the Gulf of Mexico Research Initiative and the Office of Naval Research (Grant number N00014-16-1-2192, Program Manager Dr. Reginald Beach) and is currently sponsored by the Office of Naval Research (Grant number N00014-22-1-2481, Program Manager Dr. Woei-Min Lin).  GBD's research is supported by the Office of Naval Research (Grant number N00014-17-1-2633, Program Manager Dr. Kyle Becker).
\bibliographystyle{plain}
\bibliography{mybib}

\end{multicols*}
\end{document}